\documentstyle[12pt]{article}
\begin{document}
\newcommand{\be}{\begin{equation}}
\newcommand{\ee}{\end{equation}}
\begin{titlepage}
\title{Experimental limits on antigravity \\ in extended supergravity}
\author{Stefano Bellucci \\ 
           \\{\small \it INFN--Laboratori Nazionali di Frascati,
P.O. Box 13, I-00044 Frascati, Roma (Italy)}  
\\ \\ and \\ \\
Valerio Faraoni \\ 
        \\{\small \it Department of Physics and Astronomy, University
of Victoria} \\
{\small \it P.O. Box 3055, Victoria, B.C. V8W 3P6 (Canada)}}
\date{}
\maketitle
\thispagestyle{empty}      \vspace*{1truecm}
\begin{abstract}
The available tests of the equivalence principle constrain the mass of
the Higgs-like boson appearing in extended supergravity theories.
We determine the constraints imposed by the present and future high
precision experiments on the antigravity fields arising from $N=2,8$
supergravity.
\end{abstract}
\end{titlepage}

\section{Introduction}

The discovery that $N>1$ supergravity theories lead to antigravity
is due to the work of the late J. Scherk \cite{Scherk,ScherkProc}.
In a recent paper we have revived the interest for the implications
of extended supergravity theories for antigravity \cite{belfar}.
This interest is connected to the high precision experiment at LEAR
(CERN) measuring the difference in the gravitational acceleration of
the proton and the antiproton \cite{PS-200}. For a review of earlier ideas
about antigravity
the reader is referred to the extensive article by Nieto and
Goldman \cite{GoldmanNieto} and the references therein.

The supergravity multiplet in the $N=2,8$ cases contains, in addition to
the graviton ($J=2$), a vector field $A_{\mu}^l$ ($J=1$). There
are also two Majorana gravitini ($J=\frac{3}{2}$) for $N=2$ \cite{Zachos}
and a scalar field $\sigma$ for $N=8$ \cite{Scherk,ScherkProc}.
The former fields are immaterial for our purposes and will be ignored in
the following. The field $\sigma$ does not induce any violation of the
equivalence principle. We comment below about the vanishing of the
scalar field contribution to
violations of the equivalence principle. It is also to be noted that there are
important differences between extended supergravity and the Standard Model, and
therefore the particles mentioned in this work should not be intended as the
objects familiar from the Standard Model.

The E\"otv\"os experiment forces upon us the assumption that the field
$A_{\mu}^l$ have a nonvanishing mass, which may have a dynamical origin
\cite{Scherk,ScherkProc}. In any case, the vector receives a mass
through the Higgs mechanism\setcounter{equation}{0}
\be                          \label{1}
m_l=\frac{1}{R_l}=k\, m_{\phi}\langle \phi \rangle \; ,
\ee
where the mass of the Higgs
--like
field equals its (nonvanishing) vacuum expectation value ({\em v.e.v.})
\be                          \label{2}
m_{\phi}=\langle \phi \rangle \; .
\ee
Thus, Scherk's model of antigravity leads to the possibility of
violating the equivalence principle on a range of distances of order
$R_l$, where $R_l$ is the $A_{\mu}^l$ Compton wavelength. The available
limits set by the experimental tests of the equivalence principle
allow us to constrain the {\em v.e.v.} of the scalar field $\phi$, and
therefore its mass. It must be noted that the possibility of a
massless field $A_{\mu}^l$ was already ruled out by Scherk using the
E\"{o}tv\"{o}s experiments available at that time \cite{Scherk}.

In the present paper we review the limits provided by the present day
experiments, and those obtainable from experiments currently under
planning for
the near future. The Compton wavelength of
the gravivector thus obtained is of order 10~m, or less \cite{belfar}.
Therefore,
the concept of antigravity in the context of extended supergravity
cannot play any role in astrophysics, except possibly for processes
involving the strong gravity regime, i.e. near black holes or in the
early universe.

It is worth to remind the reader that there are interesting
connections between antigravity in $N=2,$~$8$ supergravities and {\em
CP} violation experiments, via the consideration of the
$K^0$--$\overline{K}^0$ system in the terrestrial gravitational field
\cite{Scherk}. However, the present experiments on {\em CP} violations
yield bounds on the range of the gravivector field which are less
stringent than those obtained from the tests of the equivalence
principle \cite{belfar}.

The present paper has the following structure: in Sec.~2 we recall the
basic features of Scherk's antigravity. The limits on the
Compton wavelength of the gravivector and the mass of the scalar
field deriving from the current experimental verifications of the
equivalence principle are discussed in Sec.~3 . Improvements coming from
experiments under planning are also considered. The conclusions are
presented in Sec.~4.

\section{Antigravity effects in $N>1$ supergravity}

In $N=2$,~$8$ supergravity theories, the gravivector field $A_{\mu}^l$
couples to the fields of the matter scalar multiplet with strengths
\be   \label{3}
g_i=\pm k\, m_i                  \ee
\cite{Zachos} for $N=2$ and
\be
g_i=\pm 2k\,m_i                     \ee
\cite{SS,CremmerSS} for $N=8$. Here $k=(4\pi G)^{1/2}$, $m_i$ are the
quark and lepton masses, the positive and negative signs hold for
particles and antiparticles, respectively, and $g=0$ for
self--conjugated particles. As a consequence, in the interaction of an
atom with the gravitational field, the vector field $A_{\mu}^l$
``sees'' only the particles constituting the nucleon which are not
self--conjugated, while the graviton couples to the real mass of the
nucleon. The potential for an atom of atomic and mass numbers
($Z$,~$A$) in the static field of the Earth is \cite{Scherk}
\be      \label{4}
V=-\,\frac{G}{r} \left[ MM_{\oplus}-\eta M^0 {M^0_{\oplus}} \,
f\left( \frac{R_\oplus}{R_l} \right) \exp(-r/R_l) \right]  \; , \ee
where
\be    \label{5}
\eta =\left\{ \begin{array}{cllll}
1 & \,\,\,\,\,\, , & \;\;\; N=2  &  & \nonumber \\
4 & \,\,\,\,\,\, , & \;\;\; N=8  &  & \nonumber
\end{array} \right. \; .
\ee
$R_l$ is the Compton wavelength of the gravivector, and
$R_\oplus =6.38 \cdot 10^6$~m, $ M_{\oplus}=5.98 \cdot 10^{24}$~kg
are the earth radius and mass, respectively. The presence of the
function
\be     \label{6}
f(x)=3 \,\, \frac{x\cosh x-\sinh x}{x^3}      \ee
expresses the fact that a spherical mass distribution cannot be
described by a point mass located at the
center of the sphere, as in the case of a coulombic potential. The
masses in (\ref{3}) are given by
\begin{eqnarray}
&& M=Z(M_p+m_e)+(A-Z)M_n \; ,   \label{7}  \\
&& M^0=Z(2m_u+m_d+m_e)+(A-Z)(m_u+2m_d) \; ,  \label{8}
\end{eqnarray}
where $M_p$, $M_n$ and $m_e$ are the proton, neutron and electron masses,
respectively. We describe the Earth by means of the average atomic composition
$(Z_{\oplus},2Z_{\oplus})$ which gives, from (\ref{7}),~(\ref{8})
\be   \label{9}
M^0_{\oplus} \simeq \frac{3m_u+3m_d+m_e}{M_p+M_n} \, M_{\oplus} \; .
\ee
In $N=2,8$ supergravities, one of the scalar fields has a nonzero
{\em v.e.v.} and, as a consequence, the vector field $A_{\mu}^l$ acquires a
mass, as described by (\ref{2}) (the impossibility of a massless
$A_{\mu}^l$ being proved in ref.~\cite{Scherk}). This leads
to a violation of the equivalence principle, expressed by the
difference between the accelerations of two atoms with numbers
$(Z,A)$ and $(Z',A')$ in the field of the Earth
\be            \label{10}
\frac{\delta \gamma}{\gamma}=\eta \, \frac{(3m_u+3m_d+m_e)(m_e+m_u-m_d)}
{M_n \, (M_p+M_n) } \left( \frac{Z'}{A'}-\frac{Z}{A} \right) f\left(
\frac{R_{\oplus}}{R_l}\right) \left( 1+\frac{R_{\oplus}}{R_l} \right)
\exp(-R_{\oplus}/R_l) \; .                    \ee
In the spontaneously broken $N=8$ supergravity a graviscalar appears
together with the gravivector and gives a contribution to the
gravitational acceleration $\gamma $. However, this contribution has the same
sign for both particles and antiparticles, and thus cancels in the difference
$\delta \gamma $ measured in experiments on the equivalence principle. Hence,
the graviscalar does not contribute to violations of the equivalence
principle.

\section{Experimental constraints on antigravity}

In the E\"{o}tv\"{o}s--like experiment performed at the University of
Washington \cite{EotWash} (hereafter ``E\"{o}t--Wash'') the equivalence
principle was tested using berillium and copper and aluminum and
copper. The equivalence principle was verified with a precision
\be   \label{11}
\left| \frac{\delta \gamma}{\gamma} \right| < 10^{-11}  \; .
\ee
In ref.~\cite{belfar}, it was found that the Compton wavelength $R_l$ of the
gravivector satisfies the constraint
\be R_l \leq 34 \, \eta^{-1} \;\; {\mbox m} \; .
\ee
This justifies the use of the results of the E\"{o}t--Wash experiment, which
holds its validity for distances of the order $10^4$~m or less. Eq.~(\ref{11})
was used in ref.~\cite{belfar} to set a lower limit on the mass
of the Higgs--like particle:
\be    \label{12}
m_{\phi}>5 \,  \eta^{1/2} \;\;\;\;\;\mbox{GeV}        \; .    \ee
The Moscow experiment \cite{Moscow}, in spite of its higher precision,
provides a less stringent limit on $m_{\phi}$, due to the fact that it
verified the equivalence principle in the gravitational field of the
Sun, and (\ref{3}) has to be modified accordingly \cite{belfar}.
Here we make use also of the experiments aimed to detect deviations
from Newton's inverse square law, which can be seen as precise tests of the
equivalence principle. In these experiments it is customary to
parametrize the deviations from the Newtonian form with a Yukawa--like
correction to the Newtonian potential
\be   \label{13}
V(r)=-\,\frac{GM}{r} \left( 1+\alpha \, \mbox{e}^{-r/R_l} \right) \; .
\ee
In the following, we assume that, in the context
of antigravity, the parameter $\alpha$  is given by the value computed
for the E\"{o}t--Wash
experiment performed using copper ($Z=29 $, $A=63.5 $) and berillium
($Z'=4 $, $A'=9.0  $), i.e.
\be  \label{14}
\alpha =\left\{ \begin{array}{cllll}
6.36 \cdot 10^{-4} & \,\,\,\,\,\,  & \;\;\; (N=2)  &  & \nonumber \\
2.54 \cdot 10^{-3} & \,\,\,\,\,\,  & \;\;\; (N=8) \;.  &  & \nonumber
\end{array} \right.
\ee
For the materials that are likely to be used in these experiments, the
values of $\alpha$ differ from those of (\ref{14}) only for a
factor of order unity. Moreover, our final limits on $m_{\phi}$ depend
on the square root of $\alpha$. For these reasons, it is safe to use
the values (\ref{14}) of $\alpha$ in the following computations.

Equations (\ref{1}) and (\ref{2}) provide us with the relation
\be
\frac{m_{\phi}( \mbox{new})}{{m_{\phi}}^*}=\left(
\frac{{R_l}^*}{R_l( \mbox{new})}\right)^{1/2} \; ,
\ee
where ${m_{\phi}}^*=5 \eta^{1/2}$~GeV and ${R_l}^*=34 \eta^{-1}$~m are,
respectively, the 
lower limit on the scalar field mass and the 
upper limit on the Compton wavelength
of the vector $A_{\mu}^l$ derived in ref.~\cite{belfar}, and
$m_{\phi}( \mbox{new})$, $R_l( \mbox{new})$ are the new limits on the
same quantities coming from the references considered in the following.

The 2$\sigma$ limits of ref.~\cite{Spero} (see their fig.~3) allow the
range of values of $R_l$:
\be              \label{15}
R_l \leq 1 \: \mbox{cm} \:\:\:\: , \:\:\:\: R_l\geq 5 \: \mbox{cm}
\ee
for $N=2$ and
\be           \label{16}
R_l \leq 0.5 \: \mbox{cm} \:\:\:\: , \:\:\:\: R_l\geq 16 \: \mbox{cm}
\ee
for $N=8$. This corresponds to the allowed range for the mass of the
Higgs--like scalar field:
\be    \label{17}
m_{\phi} \leq 130 \: \mbox{GeV} \:\:\:\: , \:\:\:\: m_{\phi}\geq 292
\: \mbox{GeV} \:\:\:\:\:\:\:\: (N=2)
\ee
\be   \label{18}
m_{\phi} \leq 73 \: \mbox{GeV} \:\:\:\: , \:\:\:\:\: m_{\phi}\geq 412 \:
\mbox{GeV} \:\:\:\:\:\:\:\: (N=8) \; .
\ee
The curve~A of fig.~13 in ref.~\cite{Hoskinsetal} gives
\be
R_l \leq 0.6 \: \mbox{cm} \:\:\:\: , \:\:\:\: R_l\geq 10 \: \mbox{cm}
\ee
for $N=2$ and
\be
R_l \leq 0.4 \: \mbox{cm} \:\:\:\: , \:\:\:\: R_l\geq 32 \: \mbox{cm}
\ee
for $N=8$. Equivalently,
\be
m_{\phi} \leq 92 \: \mbox{GeV} \:\:\:\: , \:\:\:\: m_{\phi}\geq 376
\: \mbox{GeV} \:\:\:\:\:\:\:\: (N=2)
\ee
\be
m_{\phi} \leq 52 \: \mbox{GeV} \:\:\:\: , \:\:\:\: m_{\phi}\geq 461 \:
\mbox{GeV} \:\:\:\:\:\:\:\: (N=8) \; .
\ee
The null result of the Shternberg \cite{Shternberg} experiment reviewed by
Milyukov \cite{Milyukov} in the light of Scherk's work provides us
with the limits:
\be
R_l \leq 4 \: \mbox{cm} \:\:\:\: , \:\:\:\: R_l\geq 13 \: \mbox{cm}
\ee
for $N=2$ and
\be
R_l \leq 2.2 \: \mbox{cm} \:\:\:\: , \:\:\:\: R_l\geq 40 \: \mbox{cm}
\ee
for $N=8$. These are equivalent to:
\be
m_{\phi} \leq 82 \: \mbox{GeV} \:\:\:\: , \:\:\:\: m_{\phi}\geq 146 \:
\mbox{GeV} \:\:\:\:\:\:\:\: (N=2)
\ee
\be
m_{\phi} \leq 46 \: \mbox{GeV} \:\:\:\: , \:\:\:\: m_{\phi}\geq 197 \:
\mbox{GeV} \:\:\:\:\:\:\:\: (N=8) \; .
\ee
Therefore, the best available limits on the mass of the scalar field
are given by
\be              \label{19}
m_{\phi} \leq 82 \: \mbox{GeV} \:\:\:\: , \:\:\:\: m_{\phi}\geq
376 \: \mbox{GeV} \:\:\:\:\:\:\:\: (N=2)
\ee
\be    \label{20}
m_{\phi} \leq 46 \: \mbox{GeV} \:\:\:\: , \:\:\:\: m_{\phi}\geq 461 \:
\mbox{GeV} \:\:\:\:\:\:\:\: (N=8) \; .
\ee
A high precision test of the equivalence principle in the
field of the Earth is currently under planning in Moscow \cite{Kalebin}. The
precision expected to be achieved in this experiment is
\be     \label{21}
\left| \frac{\delta \gamma}{\gamma} \right| < 10^{-15}  \; .
\ee
Adopting the upper bound $R_l \leq 34 \, \eta^{-1}$~m found in
ref.~\cite{belfar} and approximating the function $f(x)$ for
$x=R_{\oplus}/R_l >>1$ as
\be    \label{22}
f(x) \simeq \frac{3}{2x^2} \, \mbox{e}^x \; ,
\ee
we obtain
\be      \label{23}
\frac{ \left| \delta \gamma / \gamma \right|_{\mbox{future}}}
{\left| \delta \gamma / \gamma \right|_{\mbox{E\"{o}t-Wash}}}
=10^{-4}  \; .
\ee
In the case that the new experiment verifies the equivalence principle
with the expected accuracy, the limits on $m_{\phi}$ would be pushed
to
\be  \label{24}
m_{\phi} \geq 500 \,  \eta^{1/2} \:\:\: \mbox{GeV} \; .
\ee

\section{Conclusions}

The tests of the equivalence principle and the null results on
deviations from Newton's inverse square law provide constraints on the
mass of the Higgs--like boson appearing in extended supergravity
theories. We have reviewed the limits obtainable from the available
experiments in the context of $N=2,8$ supergravity, and we have discussed
also the impact on the field of the future high precision experiment
being planned in Moscow.

There have been many papers on the effects of non--Newtonian gravity in
astrophysics, in particular those due to a fifth force like the one obtainable
from $N=2,8$ supergravity in the weak field limit (see references in
\cite{GoldmanNieto}). However, the upper bound of $34 \eta^{-1}$~m on the
Compton wavelength $R_l$ of
the gravivector field found in ref.~\cite{belfar} implies that
antigravity does not play any role in nonrelativistic astrophysics since
the length scales involved in stellar\footnote{The conclusion that stellar
structure is unaffected by antigravity might change if the non--Newtonian
force alters the equation of state of the matter composing the star
\cite{stellar}.}, galactic and supergalactic
structures dominated by gravity are much larger than $R_l$. Antigravity
could affect, in principle, processes that take place in the
strong gravity regime, where smaller distance scales are involved.
Examples of these situations are processes occurring near black hole
horizons or in the early universe, when the size of the universe is
smaller than, or of the order of, $R_l$. The relevance of antigravity
in such situations will be studied in future publications.

Our final remark concerns a point that apparently went unnoticed in
the literature on supergravity: the detection of gravitational waves
expected in a not too far future will shed light on the
correctness of supergravity theories. In fact, after the dimensional
reduction is performed, the action of the theory contains scalar and
vector fields as well as the usual metric tensor associated to spin~2
gravitons \cite{ScherkProc}. These fields are responsible for the presence of
polarization modes in gravitational waves, the effect of which
differs from that of the spin~2 modes familiar from general
relativity. Therefore, extended supergravities and general
relativity occupy different classes in the $E(2)$ classification of
Eardley {\em et al.} \cite{Eardleyetal} of gravity theories. The extra
polarization states are detectable, in principle, in a gravitational
wave experiment employing a suitable array of detectors \cite{Eardleyetal}.
However, it must be noted that a detailed study of gravitational wave
generation  taking into account the antigravity phenomenon is not available at
present. Such a work would undoubtedly have to face the remarkable
difficulties well known from the studies of gravitational wave generation in
the context of
general relativity.

\section*{Acknowledgments}

We are grateful to Prof. G. A. Lobov for drawing our attention to the ITEP
experiment. V.~F. acknowledges the warm hospitality of the INFN group in
Frascati, where this research was carried out.

\end{document}